\documentclass[
    ,final            
  ]
  {aipproc}
\usepackage{epsfig}

\layoutstyle{8x11single}

\begin{document}

\title{$\mathbf{B_s}$ Properties at the Tevatron}


\author{Sergey Burdin for the CDF and D\O\  collaborations}{
  address={Fermi National Accelerator Laboratory,
Batavia, Illinois 60510, USA} 
}

\begin{abstract}
Recent results on $B_{s}$ properties
obtained by the CDF and D\O\ collaborations using the data samples 
collected at 
the Tevatron Collider in the period 2002~--~2006 were presented
at the Hadron Collider Physics Symposium 2006 (Duke University, Durham). 
The measurements of $B_{s}$ mass and width differences are discussed in details. 
Prospects on measurements of CP violation in $B_{s}$ system are given.

\end{abstract}

\maketitle


\section{Introduction}

 Run II at the Tevatron Collider started in 2001. The CDF and D\O\ 
experiments have successfully collected data since that time. Until February 2006
each experiment recorded data corresponding to an integrated
luminosity of about 1.4~fb$^{-1}$. The analyses described in this paper
are based on samples corresponding to luminosity from 0.18 to 
$1$~fb$^{-1}$. 

 The CDF and D\O\  $B$ physics programs benefit from production of all species of $B$ hadrons 
at the Tevatron Collider.
This leads to a possibility of systematic studies of such phenomena as $B_{s}$ mixing, 
lifetime difference, rare decays and CP violation. Simultaneous measurements of relevant $B_{d}$
quantities provide very good opportunities for cross-checks of the results by comparisons with $B$ 
factories~\cite{bfac}. Though, there are more and more cases when the Tevatron $B_{d}$ results have 
comparable or even better precision. The combined $B_{s}$ and $B_{d}$ results tighten the 
overconstraint of the CKM matrix elements. Any discovered inconsistency would indicate presence of 
the new physics outside of scope of the Standard Model (SM).

 Table~\ref{tab:Results} lists the recent Tevatron $B_{s}$ results. The results on $B_{s}$ mixing, lifetime difference and CP violation 
will be discussed in details.
\begin{table}[htb]
\begin{tabular}{l|cc|cc}
\hline
  \tablehead{1}{c}{b}{Quantity}
  & \tablehead{1}{c}{b}{CDF} 
  & \tablehead{1}{c} {} {$\left(\int\mathcal{L}\mathrm{d}t,\ \mathrm{fb}^{-1}\right)$} 
  & \tablehead{1}{c}{b}{D\O\ }
  & \tablehead{1}{c} {} {$\left(\int\mathcal{L}\mathrm{d}t,\ \mathrm{fb}^{-1}\right)$} \\
\hline
$\Delta m_{s}$,~ps$^{-1}$ & $17.33^{+0.42}_{-0.21}\pm 0.07$ & (1) & $17 - 21 @ 90\%$ & (1) \\
$\Delta \Gamma_{s}$,~ps$^{-1}$ & $0.47^{+0.19}_{-0.24}\pm 0.01$ & (0.260) & $0.15\pm 0.10^{+0.03}_{-0.04}$ & (0.800) \\
$\Delta \Gamma_{CP}/\Gamma_{CP} (B_{s}\to KK)$~\cite{satoru} & $-0.08\pm 0.23\pm 0.03$ & (0.360) & --- & \\
$c\tau_{s}$,~ps & $1.381\pm 0.055^{+0.052}_{-0.046}$ & (0.360) & $1.398\pm 0.044^{+0.028}_{-0.025}$ & (0.400)  \\
$Br(B_{s}\to\mu\mu)\times 10^{7}$~\cite{krutelyov} & $<1@95\%$ & (0.780) & $<2.3@95\%$ \tablenote{expected}& (1)  \\
$Br(B_{s}\to\mu\mu\phi)$~\cite{krutelyov} & $<6.7\times 10^{-5}@95\%$ & (Run I) & $<4.1\times 10^{-6}@95\%$ & (0.450)  \\
$Br(B_{s}\to D_{s}^{(*)+}D_{s}^{(*)-})$ & --- & & $0.071 \pm 0.032^{+0.029}_{-0.025}$ & (1)  \\
$Br(B_{s}\to D_{s}^{+}D_{s}^{-})/Br(B_{d}\to D_{s}^{+}D^{-})$ & $1.67\pm0.41\pm0.47$ & (0.355) & --- &  \\
$Br(B_{s}\to\phi\phi)\times 10^{3}$ & $7.6\pm 1.3\pm 0.6$ & (0.180) & --- &  \\
$Br(B_{s}\to D_{s}^{1-}\mu^{+}\nu X)\times 10^{2}$ & --- & & $0.86\pm 0.16\pm 0.16$ & (1)  \\
$Br(B_{s}\to D_{s}3\pi )/Br(B_{d}\to D^{-}3\pi )$ & $1.14 - 1.19$ & (0.355)  & --- &  \\
$Br(B_{s}\to \psi (2S)\phi )/Br(B_{s}\to J/\psi\phi )$ & $0.52\pm 0.13\pm 0.07$ & (0.360)  & $0.58\pm 0.24\pm 0.09$  & (0.300)  \\
Observation $B_{s2}^{0*}$ & --- & & $135\pm 31 ev.$ & (1) \\ 
\hline
\end{tabular}
\caption{ Recent $B_{s}$ results from Tevatron.}
\label{tab:Results}
\end{table}
\section{$\mathbf{B_{s}}$ mixing, lifetime difference and CP violation}

Time evolution of the neutral $B-\overline{B}$ systems,  $B^{0}_{d}-\overline{B}^{0}_{d}$ and $B^{0}_{s}-\overline{B}^{0}_{s}$, 
is described by the Schr\"{o}dinger equation:
\begin{equation}
i\frac{d}{dt}\left(\begin{array}{c}
|B^{0}\rangle \\ |\overline{B}^{0}\rangle
\end{array} \right)
 =
 \left( \begin{array}{cc}
M-\frac{i\Gamma}{2} & M_{12}-\frac{i\Gamma_{12}}{2}  \\
M_{12}^{*}-\frac{i\Gamma_{12}^{*}}{2} & M-\frac{i\Gamma}{2}  \end{array} \right) 
\left(\begin{array}{c}
|B^{0}\rangle \\ |\overline{B}^{0}\rangle
\end{array} \right)
\label{eq:schrod}
\end{equation}
 The mass eigenstates do not coincide with the 
corresponding flavor states (see e.g.~\cite{breport}):
$|B_{L}\rangle = p|B^{0}\rangle + q|\overline{B}^{0}\rangle,\ \ |B_{H}\rangle = p|B^{0}\rangle - q|\overline{B}^{0}\rangle$,
 where $|p|^2+|q|^2 = 1$. 
Mass differences between 
 the $B_{d(s)}$ mass eigenstates can be expressed through off-diagonal elements of the Hamiltonian from Eq.~\ref{eq:schrod}
\begin{equation}
\Delta m = M_{H} - M_{L} \approx 2|M_{12}|.
\label{eq:dm}
\end{equation}
 Corresponding lifetime differences are 
\begin{equation}
\Delta \Gamma = \Gamma_{L} - \Gamma_{H} \approx \Delta m \Re(\Gamma_{12}/M_{12}) = 2|\Gamma_{12}|\cos\varphi,\ 
\mbox{where}\ \varphi=arg(-M_{12}/\Gamma_{12}) .
\label{eq:dgamma}
\end{equation}
The non-zero off-diagonal elements of the Hamiltonian lead to
a property of $B^{0}_{d}$ and $B^{0}_{s}$ mesons to change flavor 
and transform into their antiparticles. This phenomenon is called oscillation or mixing. 
The oscillation frequency is proportional to the mass difference $\Delta m_{d(s)}$.
The phase angle $\varphi$ connects the quantities $\Delta m$ and $\Delta\Gamma$ to the third
measurable parameter $a_{fs}=\Im(\Gamma_{12}/M_{12})=(\Delta\Gamma/\Delta m)\tan\varphi$, which 
determines CP violation in mixing. The value $\Delta m_{d}$ is very well measured with the highest  
accuracy achieved at the BABAR and BELLE experiments~\cite{bd_freq}. The value $\Delta\Gamma_{d}$ is expected to be 
small due to double Cabbibo suppression 
($\Delta\Gamma_{d}/\Gamma_{d}=(2.42\pm0.59)\times 10^{-3}$~\cite{ciuchini} to be compared with the 
experimental result from BABAR and DELPHI: $\Delta\Gamma_{d}/\Gamma_{d}=(0.9\pm3.7)\times 10^{-2}$~\cite{bd_freq}). 
The SM value $a_{fs}^{d}=-(5.0\pm1.1)\times10^{-4}$~\cite{nierste,Beneke:2003az} could be enhanced in presence of new physics up to 
0.01~\cite{breport,nierste} (updated calculations are in~\cite{Lenz:2006hd}).
 The Standard Model predictions for these parameters for $B_{s}$ system 
are following: $\Delta m_{s} \sim 20$~ps$^{-1}$~\cite{utfit,ckmfit}, 
$\Delta\Gamma_{s}/\Gamma_{s}=(7.4\pm2.4)\times 10^{-2}$~\cite{ciuchini} (more recent theoretical calculations are available in~\cite{petrov}) and 
$a_{fs}^{s}=(2.1\pm0.4)\times10^{-5}$~\cite{nierste,Beneke:2003az}.  New phenomena could influence differently the 
$B_{d}$ and $B_{s}$ systems.
 
\subsection{$\mathbf{B^{0}_{s}}-\mathbf{\overline{B}^{0}_{s}}$ mixing}

  The $\Delta m_{s}$ measurements are challenging due to high $B_{s}$ oscillation frequency. It is about 40 times higher than 
 $\Delta m_{d} = 0.508\pm0.004$~ps$^{-1}$. The corresponding period of $B_{s}$ oscillations 
($\sim100$~$\mu$m)  requires to have enough events with the proper decay length resolution of the order of 
$20-25$~$\mu$m to resolve these oscillations. Significance of the oscillation signal can be expressed using the following 
formula~\cite{pdg}:
\begin{equation}
\mathcal{S}\sim\sqrt{\frac{S\varepsilon\mathcal{D}^{2}}{2}}\cdot 
\exp\left(\frac{\Delta m_{s}^{2}}{2}\left(\frac{m_{B}}{\langle p\rangle}\sigma_{L}^{2}+\left(t\frac{\sigma_{p}}{p}\right)^{2}\right)\right)\sqrt{\frac{S}{S+B}},
\label{eq:signif}
\end{equation}
where $S\ (B)$ is the number of signal (background) candidates; 
$\varepsilon$ is the tagging efficiency; $\mathcal{D}$ is the tagging dilution; $\sigma_{L}$ is
the decay length resolution;  $\sigma_{p}/p$ is the relative momentum resolution. The 
tagging dilution is related to the mistag probability $\eta$: $\mathcal{D}=1-2\eta$. Here, the tagging means determination of  
$B_{s}$ flavor at the production time. 

\begin{figure}
\begin{minipage}{0.47\textwidth}
  \epsfig{figure=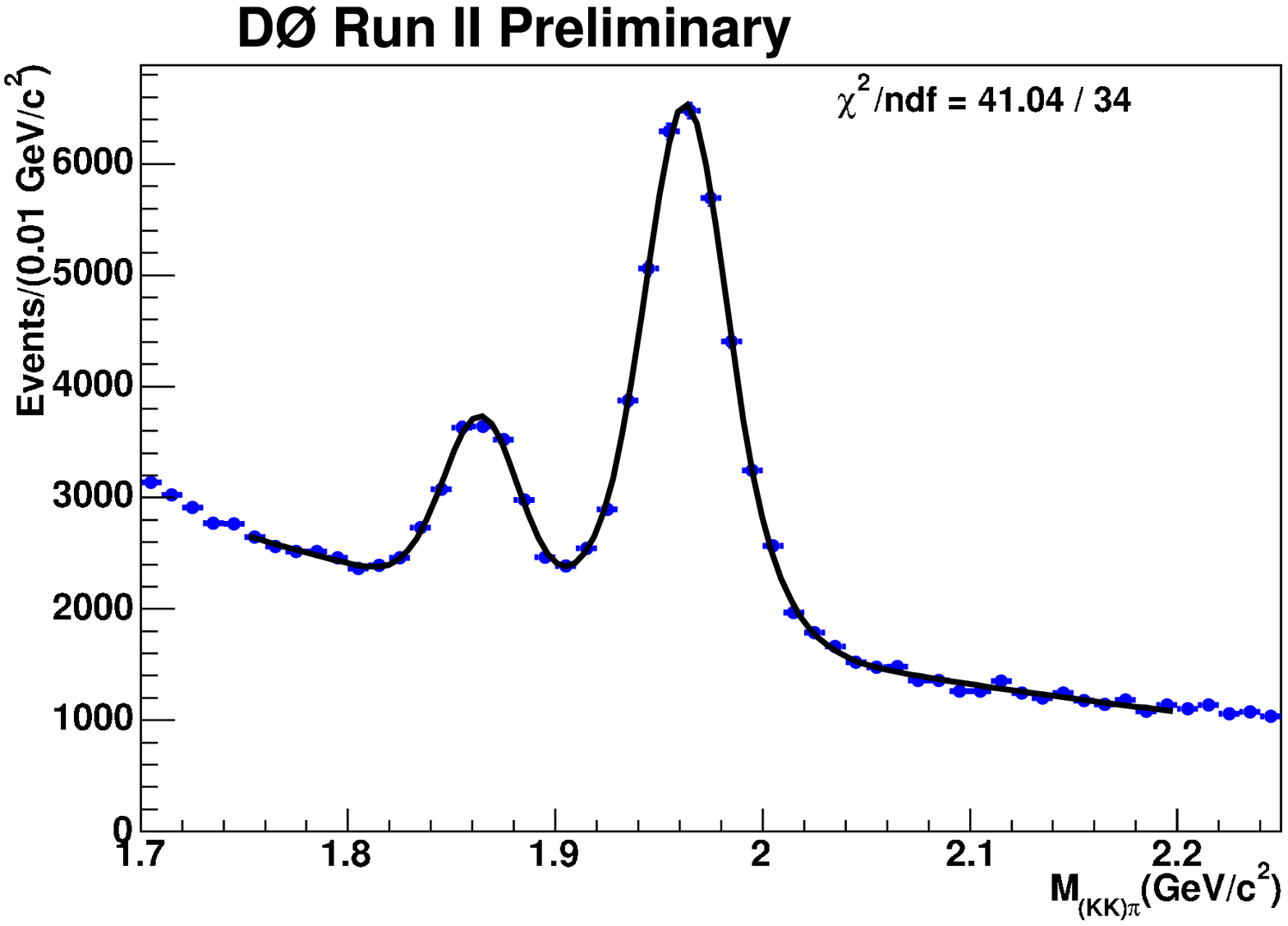,width=0.99\textwidth,height=0.3\textheight}
\end{minipage}
\hfill
\begin{minipage}{0.47\textwidth}
  \epsfig{figure=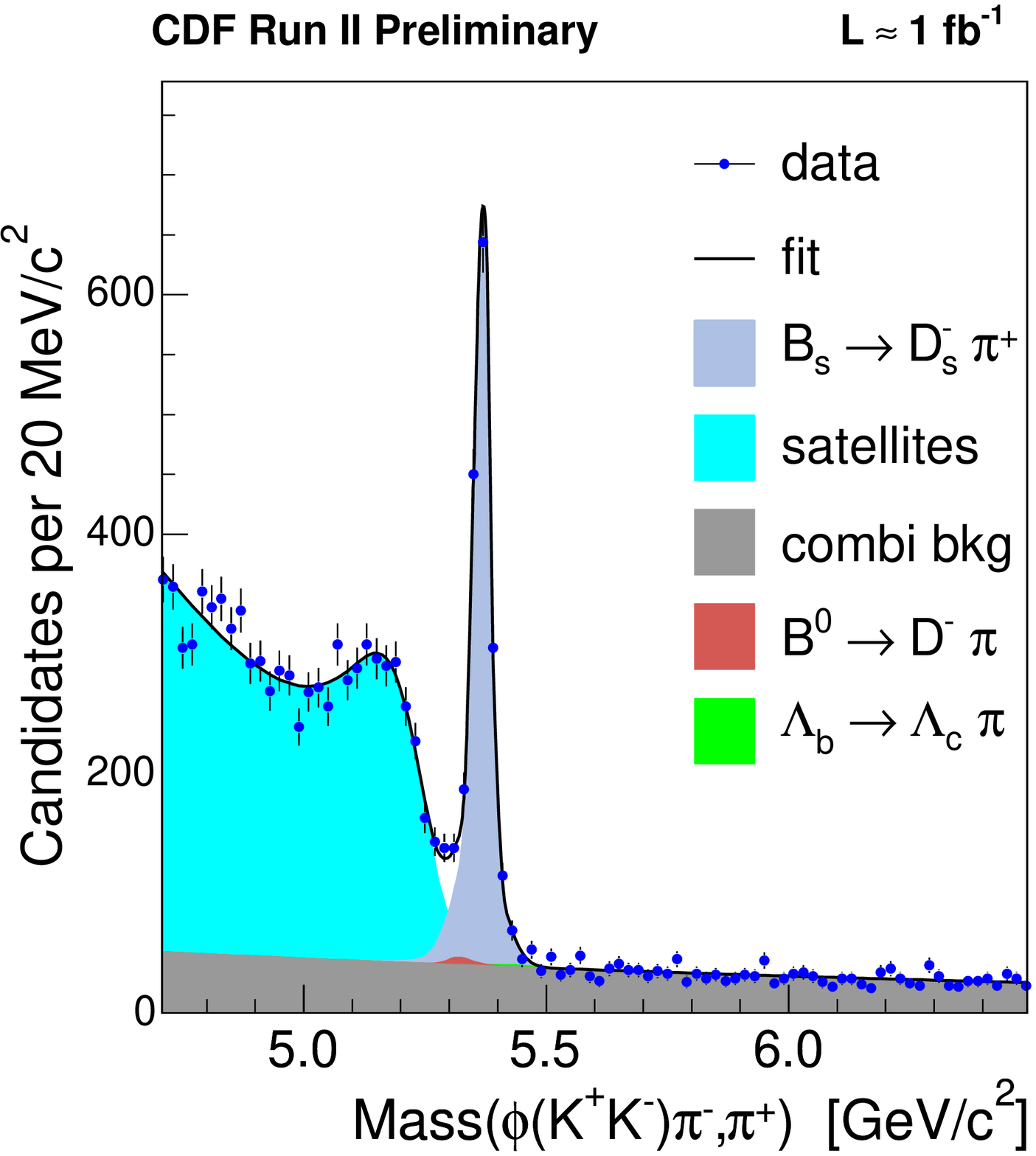,width=0.99\textwidth,height=0.3\textheight}
\end{minipage}
\caption{$B_{s}^{0}$ signal samples at D\O\  (left) and CDF (right).}
\label{fig:d0_cdf_bs_signal}
\end{figure}

  Both CDF and D\O\  used data samples corresponding to $1$~fb$^{-1}$ of integrated luminosity in the 
$B_{s}$ oscillation analyses. 
The CDF strategy for collecting the $B_{s}$ samples is based on the displaced track triggers and D\O\  exploited its muon 
system. The D\O\  experiment collected $26,710\pm556$ $B_{s}\to X\mu\nu D_{s}(\to\phi\pi)$  candidates shown in 
Fig.~\ref{fig:d0_cdf_bs_signal} (left).  
CDF reconstructed $3,600$ hadronic $B_{s}^{0}\to D_{s}^{-}(\pi^{+}\pi^{-})\pi^{+}$ 
and $37,000$ semileptonic $B_{s}^{0}\to l^{+}D_{s}^{-}X\ (l=e,\mu)$ decays. In both cases the modes 
$D_{s}^{-}\to \phi\pi^{-},\ K^{*0}K^{-},\ \pi^{+}\pi^{-}\pi^{-}$ were used.
The hadronic $B_{s}^{0}\to D_{s}^{-}\pi^{+}$ 
sample is shown in Fig.~\ref{fig:d0_cdf_bs_signal} (right). Semileptonic  decays have much broader distribution on 
reconstructed $B_{s}$ momentum resolution ($3-20\%$) in comparison with fully reconstructed hadronic decays. 
Equation~\ref{eq:signif} shows that this resolution becomes important for large proper decay times. This decreases 
significantly power of the semileptonic $B_{s}$ samples. 

  Calibration of the decay length resolution is essential for the $B_{s}$ mixing analyses due to high $\Delta m_{s}$ 
oscillation frequency. CDF utilized large sample of prompt $D^{+}$ mesons combined with one or three tracks from
the primary vertex. This combination effectively simulates the $B_{s}^{0}\to D^{-}_{s}(\pi^{+}\pi^{-})\pi^{+}$
topology with known ``$B_{s}^{0}$'' decay vertex allowing to calibrate the vertex resolution. D\O\  used 
$J/\psi\to \mu^{+}\mu^{-}$ sample where $\sim 70\%$ of $J/\psi$ mesons are prompt. Overall
decay length resolution scale factors have been 
determined using this sample: $1.0$ for $72\%$ of events and $1.8$ for  the rest. 
Simulated events were used to check a dependence of these scale factors from events topologies.

 The tagging  utilizes information from fragmentation track at the $B_{s}$ reconstruction side 
(same-side tagging) or tries to determine the $B$ flavor at the opposite side through 
partial reconstruction of its decay products (opposite-side tagging). The first technique is 
characterized by high efficiency $\varepsilon$ and relatively low dilution $\mathcal{D}$. 
The opposite-side tagging has 
low efficiency but higher dilution. As can be seen from equation~\ref{eq:signif} the tagging power
is determined by combination of these two parameters: $\varepsilon\mathcal{D}^2$. The opposite-side 
tagging was calibrated using $B_{d}$ and $B_{u}$ samples. The opposite-side
tagging power was measured to be equal $\varepsilon\mathcal{D}^2 = 2.5\pm 0.2\%$ at D\O\  and 
 $\varepsilon\mathcal{D}^2 = 1.5\pm 0.1\%$ at CDF. The same-side tagging was used at CDF with the power 
 $\varepsilon\mathcal{D}^2 = 3.5\%\ (4.0\%)$ for the hadronic (semileptonic) sample
determined using the PYTHIA Monte Carlo simulated events. Particle identification used for selection of the 
fragmentation track significantly improved the same-side tagging power.

\begin{figure}
\begin{minipage}{0.47\textwidth}
  \epsfig{figure=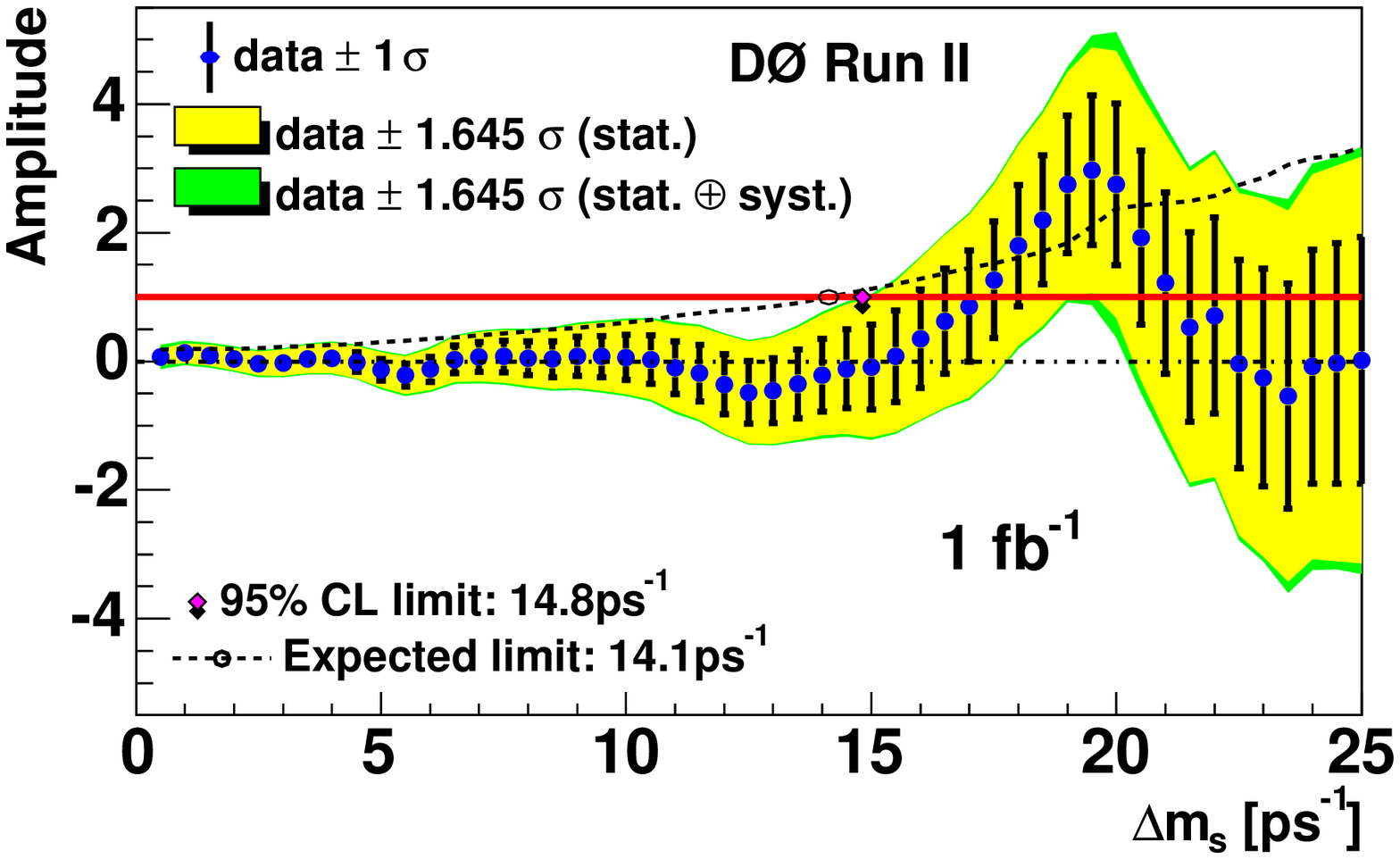,width=0.99\textwidth,height=0.3\textheight}
\end{minipage}
\hfill
\begin{minipage}{0.47\textwidth}
  \epsfig{figure=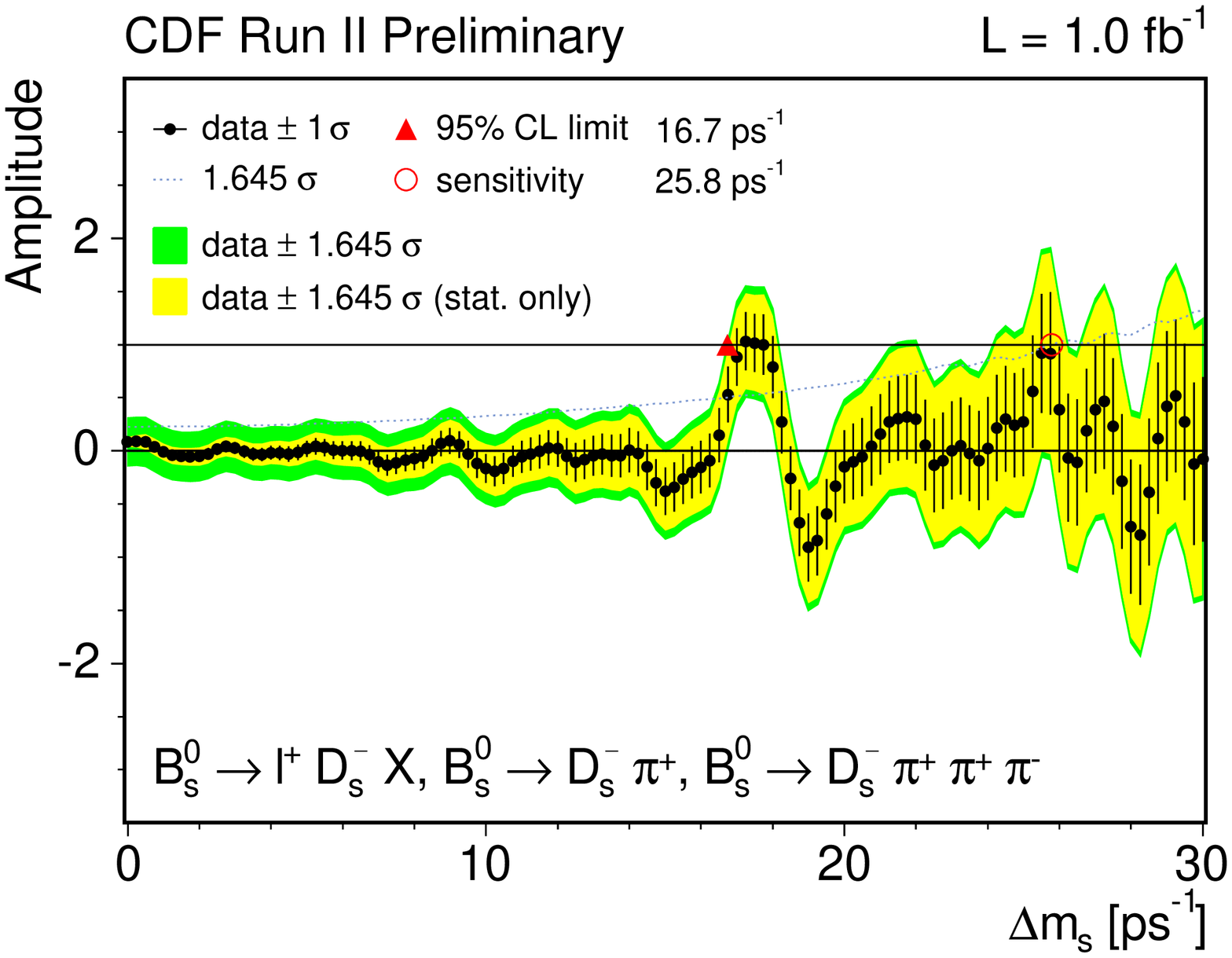,width=0.99\textwidth,height=0.3\textheight}
\end{minipage}
\caption{$B_{s}^{0}$ amplitude scan at D\O\   (left) and CDF (right).}
\label{fig:d0_cdf_bs_ampl}
\end{figure}

  Probability for a $B_{s}$ candidate to be reconstructed as oscillated (changed flavor with respect to the production time) or 
non-oscillated is following:
\begin{equation}
p_{s}^{nos/osc} = \frac{K}{2\tau_{B_{s}}}e^{-\frac{Kx}{c\tau_{B_{s}}}}\left(1\pm\mathcal{D}\cos(\Delta m_{s}\cdot Kx/c)\right),\ \mbox{where}\ K=\frac{p_{\mu D_{s}}}{p_{B_{s}}}. 
\label{eq:prob}
\end{equation}
  To detect a signal the amplitude scan method is used~\cite{moser}. The probability is modified adding the parameter called amplitude
$\mathcal{A}$ to the cosine term: $\cos(\Delta m_{s}\cdot Kx/c)\cdot\mathcal{A}$.
  The amplitude $\mathcal{A}$ is consistent with $1$ for  $\Delta m_{s} = \Delta m_{s}^{true}$ and otherwise consistent
with $0$. Fig.~\ref{fig:d0_cdf_bs_ampl} shows the amplitude scans from D\O\   (left) and CDF (right). The D\O\  amplitude scan 
shows $2.5\sigma$ deviation from $0$ at $19$~ps$^{-1}$ with the expected $95\%$ CL limit $14.1$~ps$^{-1}$. The 
CDF amplitude scan reveals the signal around $17$~ps$^{-1}$ with the expected $95\%$ CL limit $25.3$~ps$^{-1}$. 

\begin{figure}
\begin{minipage}{0.47\textwidth}
  \epsfig{figure=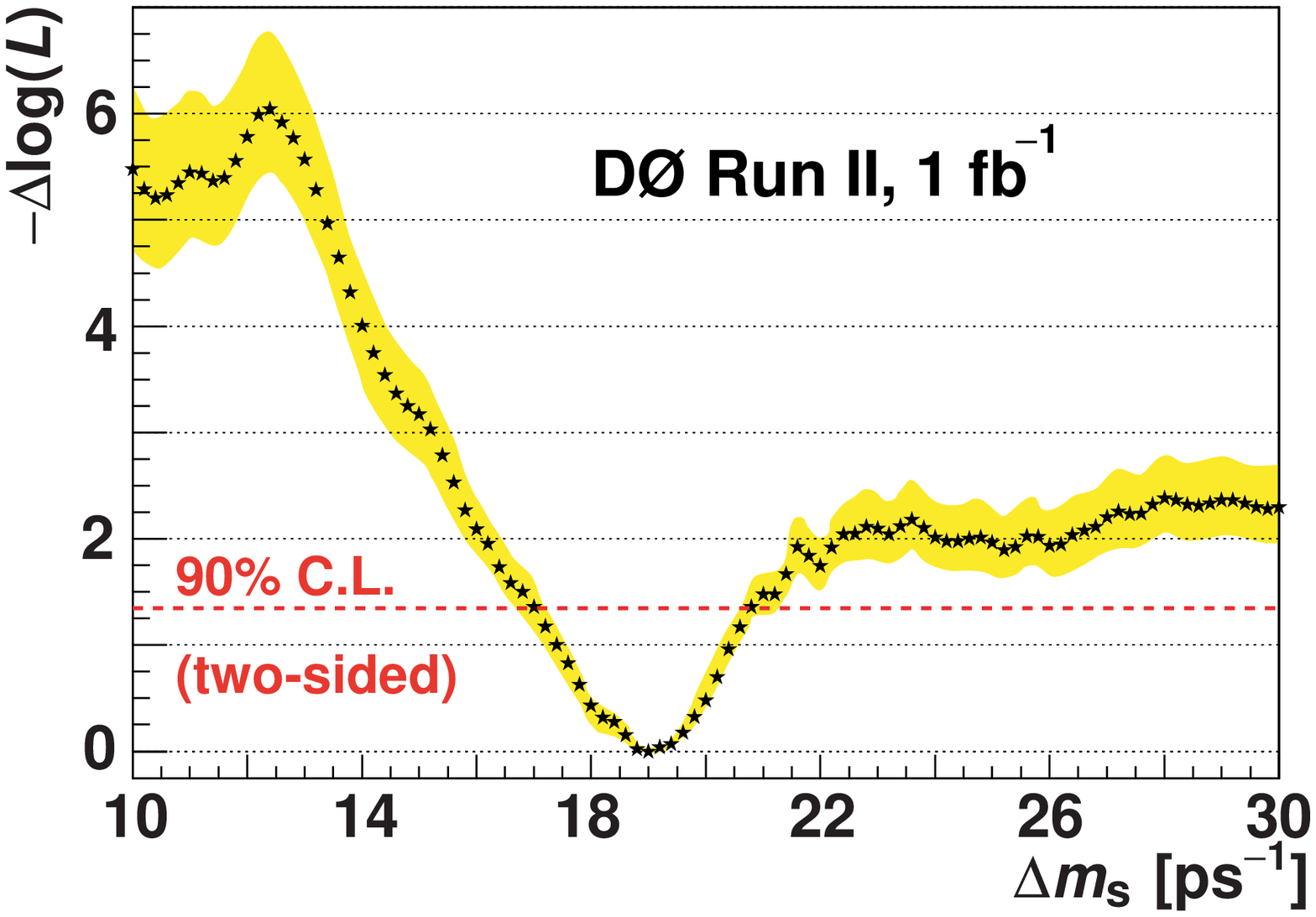,width=0.99\textwidth,height=0.3\textheight}
\end{minipage}
\hfill
\begin{minipage}{0.47\textwidth}
  \epsfig{figure=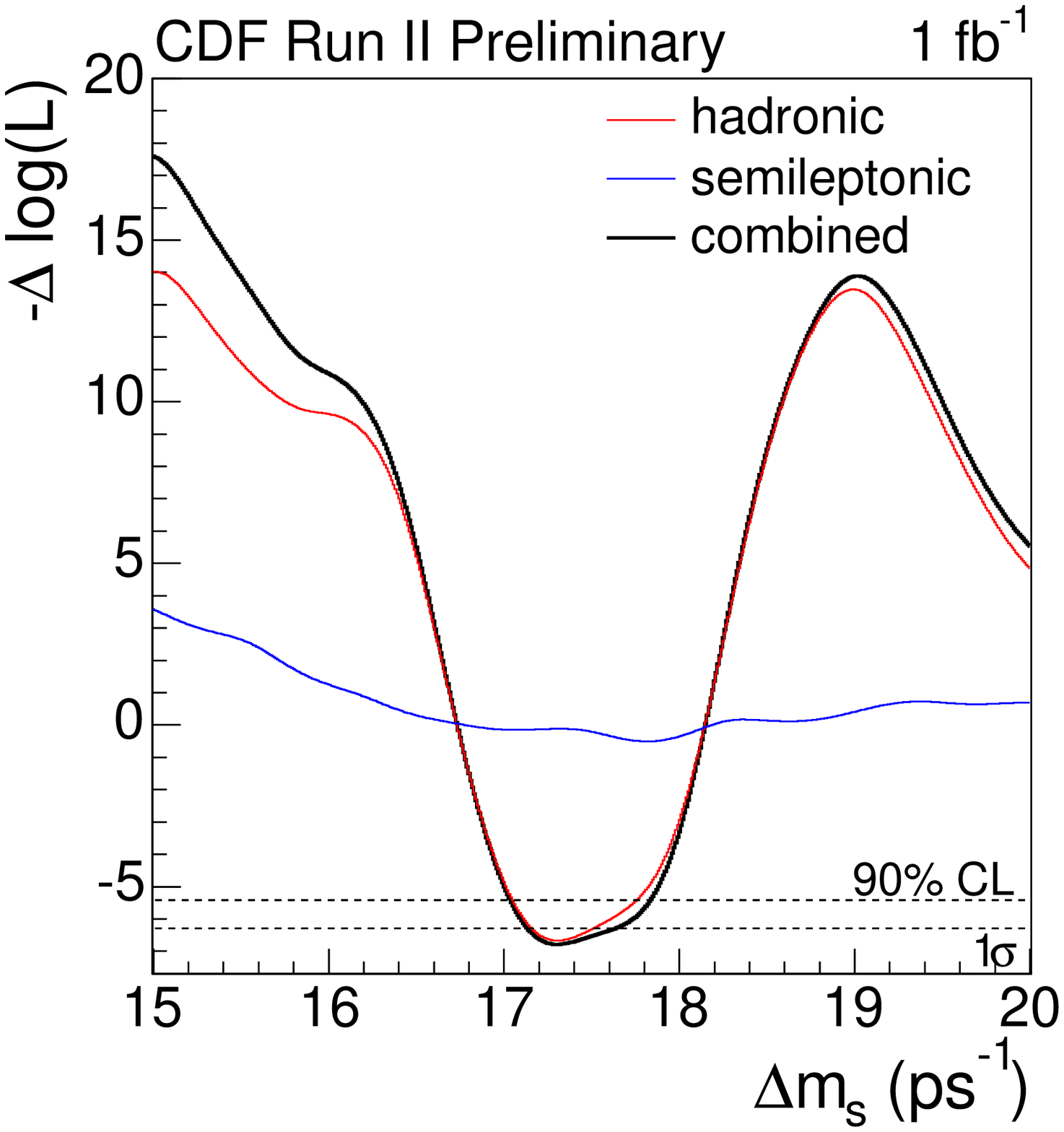,width=0.99\textwidth,height=0.3\textheight}
\end{minipage}
\caption{$B_{s}^{0}$ log likelihood scan at D\O\  (left) and CDF (right).}
\label{fig:d0_cdf_bs_logl}
\end{figure}

  The log likelihood scans (Fig.~\ref{fig:d0_cdf_bs_logl}) are in agreement with the amplitude scans. D\O\  sets the two-sided 
limit $17<\Delta m_{s}< 21$ at $90\%$ CL~\cite{d0_bsmix}. The probability of background fluctuation to give signal of the same significance
is $5\%$. The corresponding CDF result is $17.01<\Delta m_{s}< 17.84$~ps$^{-1}$ at $90\%$ CL with the probability of 
background fluctuation $0.2\%$~\cite{cdf_bsmix}. The central value of $B_{s}$ oscillation frequency from CDF is 
$\Delta m_{s}=17.31^{+0.33}_{-0.18}(\mbox{stat.}\pm0.07(\mbox{syst.})$~ps$^{-1}$ in good agreement 
with the theoretical SM predictions. 

\subsection{Lifetime difference in $\mathbf{B^{0}_{s}}-\mathbf{\overline{B}^{0}_{s}}$ system}

  Measurements of the lifetime difference in $B^{0}_{s}-\overline{B}^{0}_{s}$ system is possible through study
of the $B_{s}$ decays with common final states for $B^{0}_{s}$ and $\overline{B}^{0}_{s}$. Examples of such final 
states are $J/\psi\phi$, $D^{(*)+}_{s}D^{(*)-}_{s}$ and $K^{+}K^{-}$ (theoretical calculations are given 
in~\cite{Fleischer:1999,Fleischer:2001,Fleischer:0705.1121}).
The Tevatron presented results on all these decays.

The decay $B_{s}\to D^{+}_{s} D^{-}_{s}$ has pure CP-even final state. It is expected that the  
inclusive decays $B_{s}\to D^{(*)+}_{s} D^{(*)-}_{s}$ is also CP-even with $5\%$ uncertainty. 
Then, a measurement of the branching ratio $Br\left(B_{s}\to D^{(*)+}_{s} D^{(*)-}_{s}\right)$ leads to
$\Delta\Gamma_{CP}$:
\begin{equation}
\frac{\Delta\Gamma_{CP}}{\Gamma}\sim \frac{2Br(B_{s}\to D^{(*)+}_{s} D^{(*)-}_{s})}{1-Br(B_{s}\to D^{(*)+}_{s} D^{(*)-}_{s})/2}.
\label{eq:dgamma_cp}
\end{equation}
$\Delta\Gamma_{CP}$ is equal to $\Delta\Gamma_{s}$ assuming $\varphi=0$ (see Eq.~\ref{eq:dgamma}).

\begin{figure}
\begin{minipage}{0.47\textwidth}
  \epsfig{figure=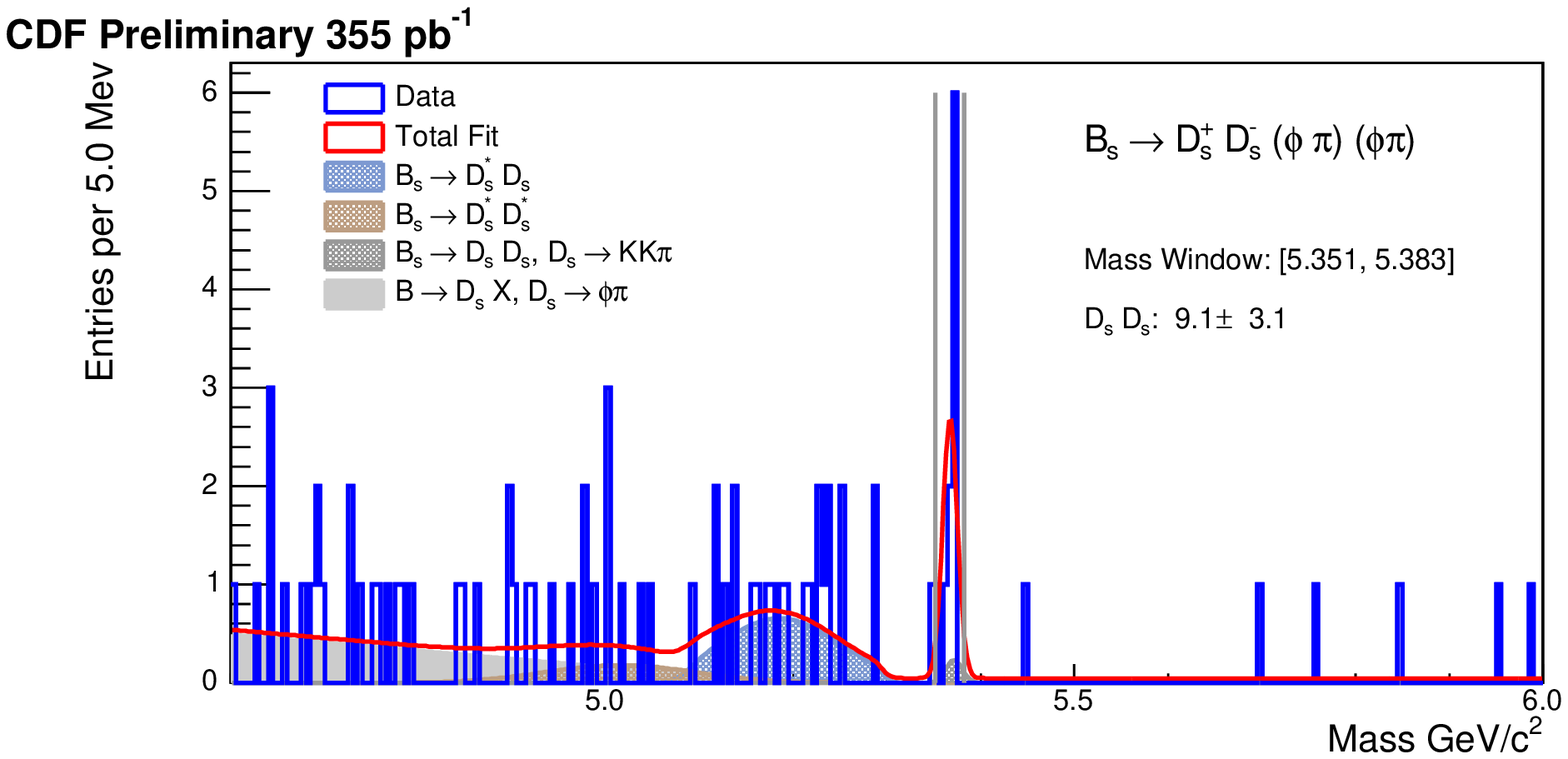,width=0.99\textwidth,height=0.3\textheight}
\end{minipage}
\hfill
\begin{minipage}{0.47\textwidth}
  \epsfig{figure=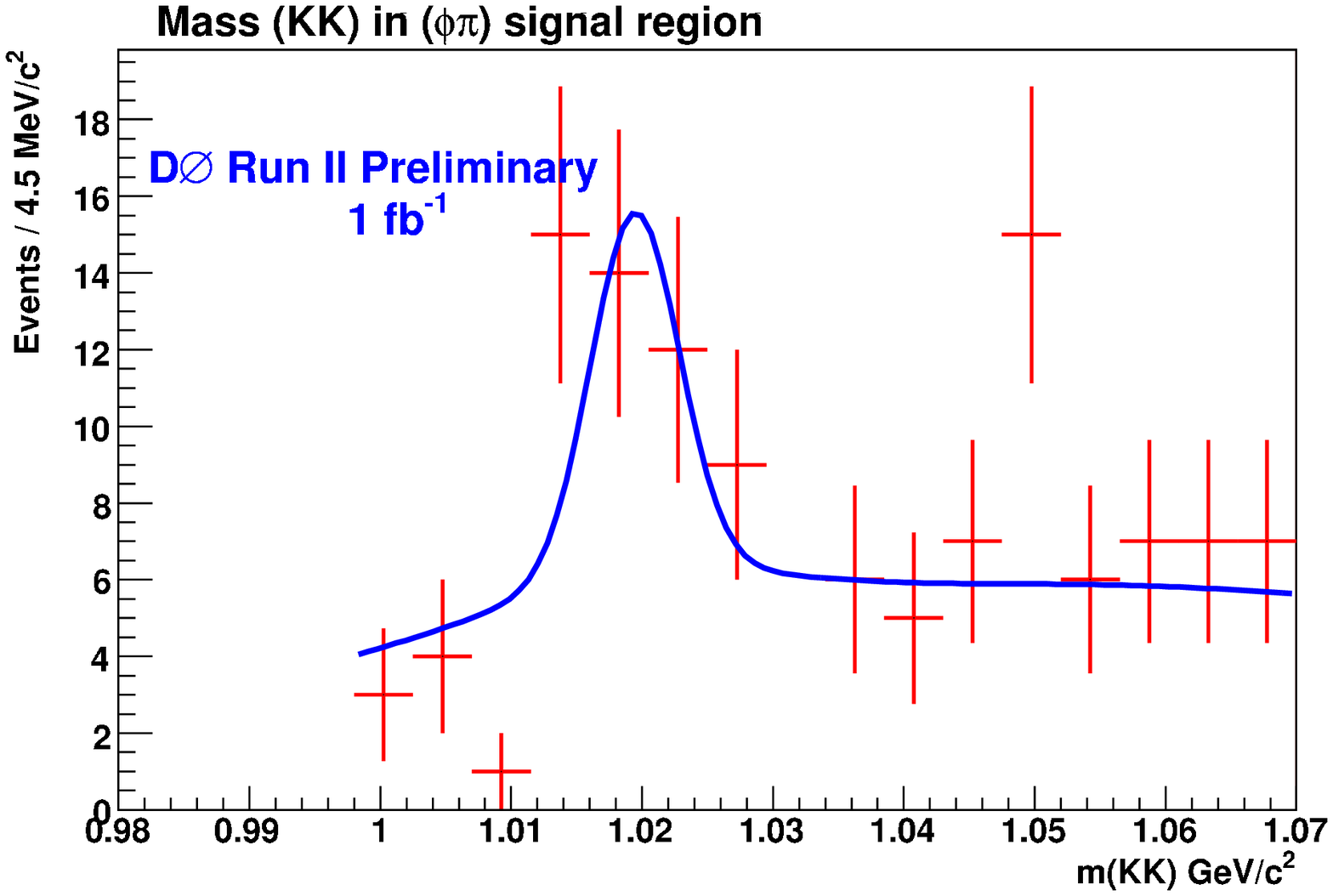,width=0.99\textwidth,height=0.3\textheight}
\end{minipage}
\caption{$B_{s}^{0}\to D^{(*)+}_{s} D^{(*)-}_{s}$ signal at CDF (left) and D\O\  (right).}
\label{fig:d0_cdf_dsds}
\end{figure}

  CDF reconstructed $23.5\pm5.5$ candidates of 
the decay $B_{s}\to D^{+}_{s}(\to\phi\pi^{+}) D^{-}_{s}(\to\phi\pi^{-})$ (see Fig.~\ref{fig:d0_cdf_dsds} (left)). The 
branching ratio was measured relative to the decay $B_{d}\to D^{+}_{s} D^{-}$:  
$Br(B_{s}\to D_{s}^{+}D_{s}^{-})/Br(B_{d}\to D_{s}^{+}D^{-})=1.67\pm0.41\pm0.47$~\cite{cdf_dsds}. 
Work on $\Delta \Gamma_{CP}$ 
measurement is in progress.

  D\O\  used semileptonic $D_{s}$ decays due to trigger requirements and reconstructed $19.3\pm7.8$ candidates of the decay
$B_{s}\to D^{(*)+}_{s}(\to\phi\mu^{+}\nu) D^{(*)-}_{s}(\to\phi\pi^{-})$ (see Fig.~\ref{fig:d0_cdf_dsds} (right)). 
As a normalization process the decay
$B_{s}\to D^{(*)+}_{s}(\to\phi\pi^{+})\mu^{+}\nu$ was chosen. The branching ratio
$Br(B_{s}\to D^{(*)+}_{s} D^{(*)-}_{s})=0.071\pm0.032(\mbox{stat.})^{+0.029}_{-0.025}(\mbox{syst.})$ was measured. Using 
Eq.~\ref{eq:dgamma_cp} the value 
$\Delta\Gamma_{CP}/\overline{\Gamma}_{s}=0.142\pm0.064(\mbox{stat.})^{+0.058}_{-0.050}(\mbox{syst.})$
was determined~\cite{d0_dsds}. 

 CDF determined the lifetime difference
$\Delta \Gamma_{CP}(B_{s}\to K^{+}K^{-})/\Gamma_{CP} (B_{s}\to K^{+}K^{-})=-0.08\pm 0.23\pm 0.03$~\cite{cdf_kk}
using the  $B_{s}$ lifetime measurement in the $K^{+}K^{-}$ final state:
$\tau(B_{s}\to K^{+}K^{-})=1.53\pm 0.18 (stat.)\pm 0.02 (syst.)$~ps~\cite{satoru}.

\begin{figure}
\begin{minipage}{0.47\textwidth}
  \epsfig{figure=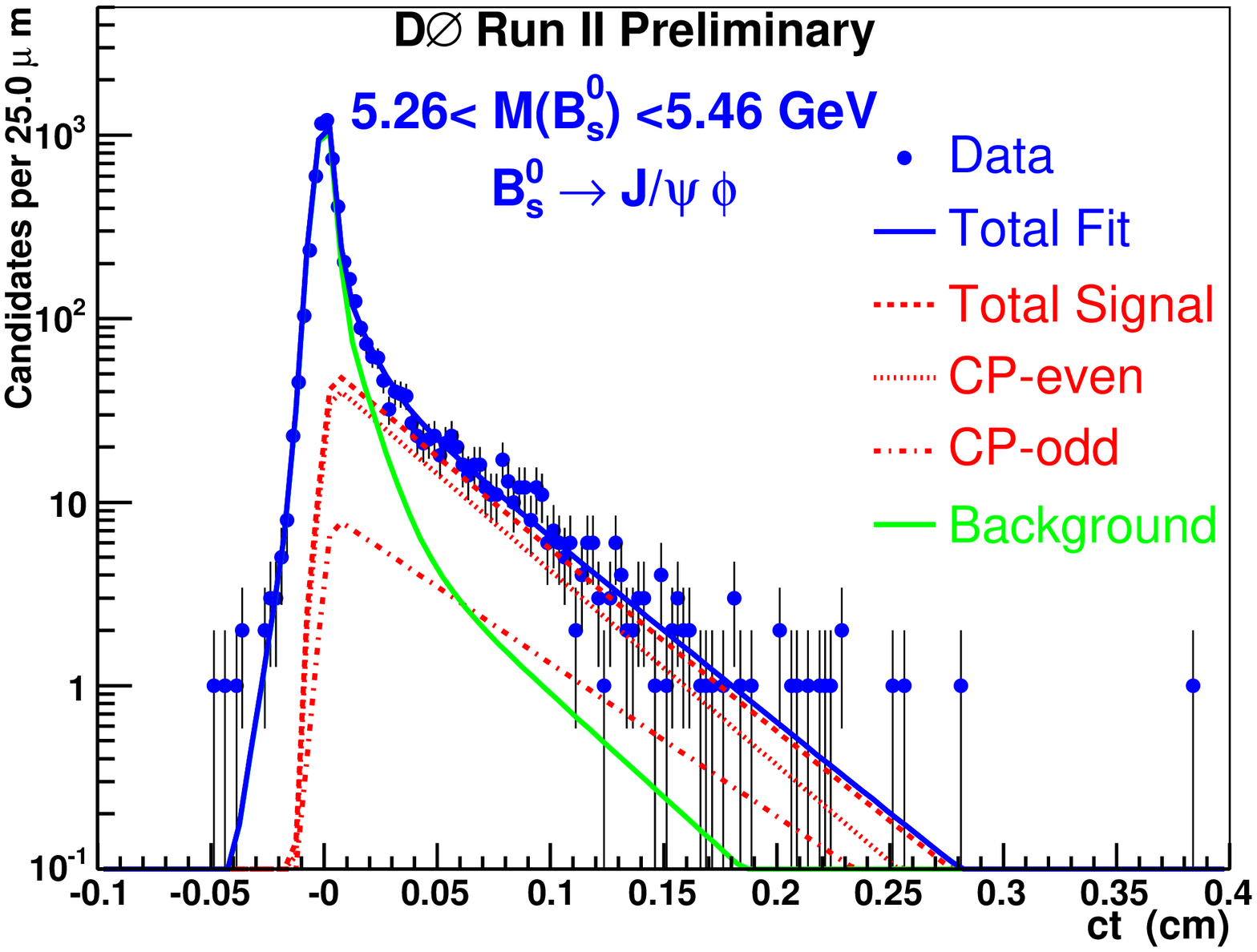,width=0.99\textwidth,height=0.3\textheight}
\end{minipage}
\hfill
\begin{minipage}{0.47\textwidth}
  \epsfig{figure=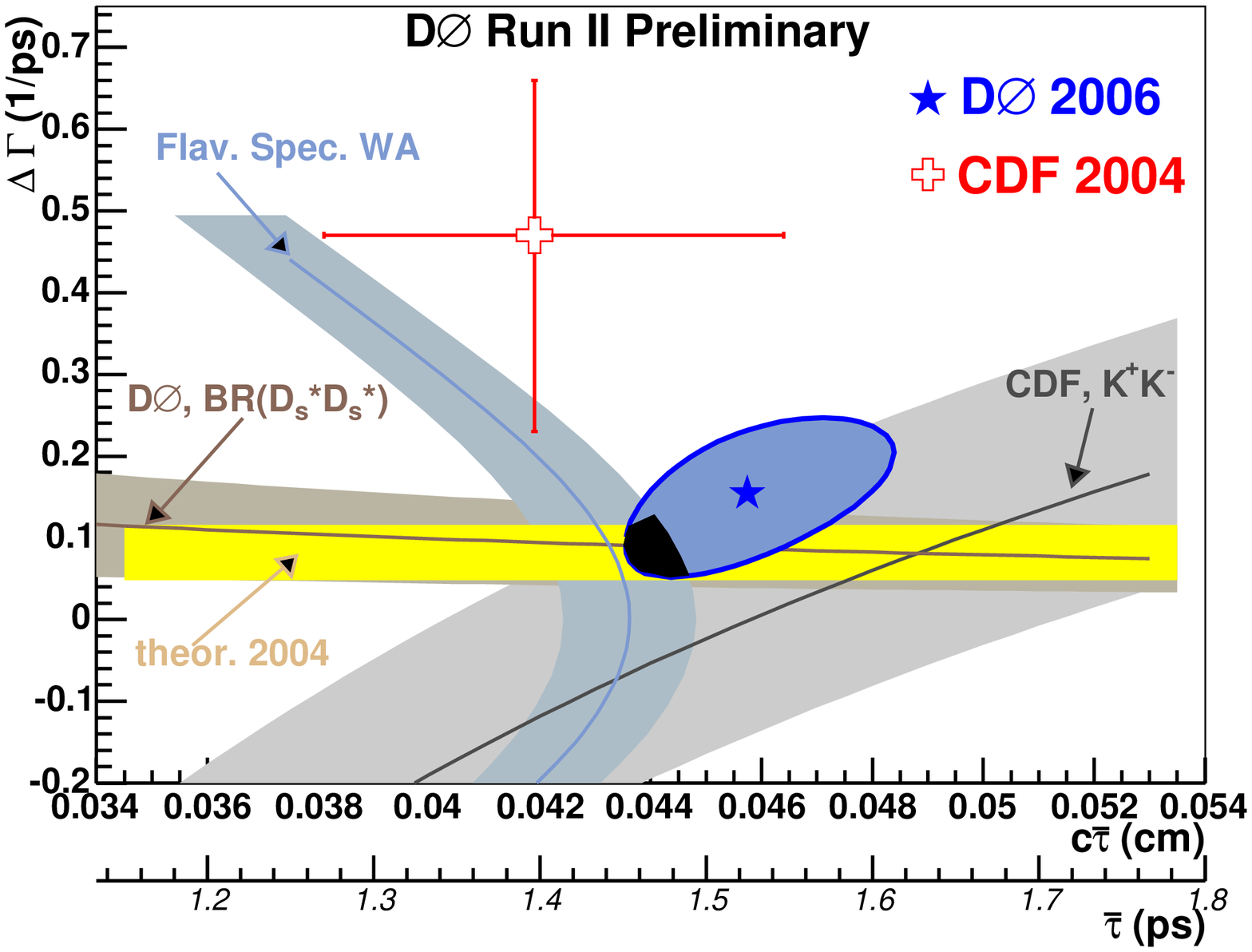,width=0.99\textwidth,height=0.3\textheight}
\end{minipage}
\caption{$B_{s}$ CP-even and CP-odd lifetimes from D\O\  (left). 
Dependence of $\Delta\Gamma_{s}$  results from average $B_{s}$ lifetime (right). ``CDF 2004'' and ``D\O\  2006'' results refer to 
the $B_{s}\to J/\psi\phi$ analyses.}
\label{fig:d0_dg_comb}
\end{figure}

  The final state  $J/\psi\phi$ is a mix of CP-even and CP-odd states which can be separated using angular distributions and 
the corresponding lifetimes can be measured (Fig.~\ref{fig:d0_dg_comb} (left)). The D\O\   result updated using $0.8$~fb$^{-1}$ is 
$\Delta\Gamma_{s} = 0.15\pm0.10(stat.)^{+0.03}_{-0.04}(syst.)$~\cite{d0_dgam}.

 Fig.~\ref{fig:d0_dg_comb} shows the $\Delta\Gamma_{s}$ results as functions of average $B_{s}$ lifetime. 
The SM theoretical prediction~\cite{beneke} is shown as the horizontal band.

\subsection{CP violation}

  The Tevatron experiments have possibilities to measure both direct CP violation and CP violation in mixing. 

  The direct CP violation can be measured using the decay $B_{s}^{0}\to K^{-}\pi^{+}$~\cite{Fleischer:0705.1121}. CDF collected a sample
of hadronic two-body B decays which consists of 
$B^{0}_{d}\to \pi^{+}\pi^{-}$, $B^{0}_{d}\to K^{+}\pi^{-}$,
$B^{0}_{s}\to K^{+}K^{-}$ and $B^{0}_{s}\to K^{-}\pi^{+}$. The measurement of CP violation using this sample has good accuracy 
and compatible with B-factories~\cite{belle,babar}: 
$A_{CP}^{CDF}(B^{0}_{d}\to K^{+}\pi^{-})=-0.058\pm0.039(\mbox{stat.}\pm0.007(\mbox{syst.}))$~\cite{cdf_kpi}. 
The next step is an observation of $B^{0}_{s}\to K^{-}\pi^{+}$ decay and 
determination of the direct CP violation in the $B_{s}$ system which could be a model-independent 
probe for new phenomena~\cite{lipkin,Fleischer:0705.1121}.

  D\O\  obtained the world most precise result on the CP violation in mixing in $B_{d}$ system: 
$\Re(\varepsilon_{B})/(1+|\varepsilon_{B}|^{2})=a_{fs}^{d}/4=-(1.1\pm1.0\pm0.7)\times 10^{-3}$~\cite{d0_cpv}. 
Changes in the magnet polarities during 
different periods of data taking help to reduce systematic uncertainties in the CP violation measurements. This work was an
important step toward the CP violation in mixing measurement in $B_{s}$ system~\cite{cpviol_bs}.

\section{Conclusion}

 Complex studies of the $B_{s}$ properties are being conducted using the CDF and D\O\  detectors at
the Tevatron Collider. The results on $B_{s}$ mixing, lifetime difference and first steps toward the CP 
violation measurements in $B_{s}$ system were discussed in details.

\begin{theacknowledgments}
  The author thanks the organizers of the Symposium for very interesting program and the physics 
analysis representatives from CDF and D\O\   for providing results.
\end{theacknowledgments}

\bibliographystyle{aipprocl} 


\end{document}